\documentclass[Crown,sagev,times,10pt, doublespace]{sagej}

\usepackage{moreverb, url}
\usepackage[colorlinks,bookmarksopen,bookmarksnumbered,citecolor=red,urlcolor=red]{hyperref}
\usepackage{subfigure}
\usepackage{multirow}
\usepackage{amssymb} 
\usepackage{array}
\usepackage{geometry}

\newcommand\BibTeX{{\rmfamily B\kern-.05em \textsc{i\kern-.025em b}\kern-.08em
T\kern-.1667em\lower.7ex\hbox{E}\kern-.125emX}}

\def\volumeyear{2024}

\begin{document}

\runninghead{Zhang et al.}

\title{Statistical Considerations for Evaluating Treatment Effect under Various Non-proportional Hazard Scenarios}

\author{Xinyu Zhang\affilnum{1}, Erich J. Greene\affilnum{1}, Ondrej Blaha*\affilnum{1} and Wei Wei*\affilnum{2}}

\affiliation{\affilnum{1}Department of Biostatistics, Yale School of Public Health, CT\\
\affilnum{2}Department of Internal Medicine, Yale School of Medicine, CT}

\corrauth{Wei Wei, Department of Internal Medicine, Yale School of Medicine, CT \\ Ondrej Blaha, Department of Biostatistics, Yale School of Public Health, CT}
\email{wei.wei@yale.edu;  ondrej.blaha@yale.edu}

\begin{abstract}
We conducted a systematic comparison of statistical methods used for the analysis of time-to-event outcomes under various proportional and nonproportional hazard (NPH) scenarios. Our study used data from recently published oncology trials to compare the Log-rank test, still by far the most widely used option, against some available alternatives, including the MaxCombo test, the Restricted Mean Survival Time Difference ($\Delta$RMST) test, the Generalized Gamma Model (GGM) and the Generalized F Model (GFM). Power, type I error rate, and time-dependent bias with respect to the RMST difference, survival probability difference, and median survival time were used to evaluate and compare the performance of these methods.
In addition to the real data, we simulated three hypothetical scenarios with crossing hazards chosen so that the early and late effects “cancel out” and used them to evaluate the ability of the aforementioned methods to detect time-specific and overall treatment effects. We implemented novel metrics for assessing the time-dependent bias in treatment effect estimates to provide a more comprehensive evaluation in NPH scenarios.
Recommendations under each NPH scenario are provided by examining the type I error rate, power, and time-dependent bias associated with each statistical approach.
\end{abstract}


\maketitle

\section{Introduction}
The Log-rank test is commonly used to test the efficacy of a novel treatment against a standard of care in studies involving time-to-event outcomes. Under the proportional hazards (PH) assumption, the treatment effect between control and treatment arms can be summarized by the hazard ratio (HR), with a value of less than one representing a ``positive'' treatment effect and a value larger than one representing a ``negative'' effect in any situation when the outcome of interest is, for example, time to death or disease progression. If the two groups have the same survival distribution, the Log-rank statistic approximately follows a standard normal distribution \cite{schoenfeld1981asymptotic} and is asymptotically equivalent to the likelihood ratio test statistic for any family of distributions with a proportional hazard alternative. However, it is well known that the statistical power of the Log-rank test can be compromised when the PH assumption is violated. This is of particular concern in immuno-oncology trials, where the treatment effect is very likely to be time-varying rather than constant over time. Furthermore, interpretation of the hazard ratio, which is commonly used for summarizing and quantifying the effect of a treatment, becomes difficult when the PH assumption is severely violated.   

Immuno-oncology (IO) is a rapidly evolving area in anticancer drug development. Multiple high-profile IO trials \cite{HodiIO, wolchok2010ipilimumab, small2006placebo} have shown evidence of possible delayed treatment effects, a typical nonproportional hazard (NPH) pattern where the two survival curves overlap initially but start to separate after some period of time. Several underlying reasons may contribute to a delayed treatment effect,\cite{alexander2018hazards} such as the presence of heterogeneous subpopulations,\cite{rahman2019divining} the trial design, \cite{disis2014mechanism} or the unique mechanism of action of the treatment. Another common type of NPH pattern is a crossing pattern, where the survival functions of the two trial arms cross each other at some point during the study follow-up; in the literature, such patterns have arisen due to a subgroup of patients developing treatment-resistant mutations\cite{mok2009gefitinib} and due to a treatment being harmful in a specific subgroup of patients while being beneficial for the rest of the enrolled participants.\cite{ananthakrishnan2021critical}
Heterogeneous subpopulations can also contribute to the distance between survival curves diminishing over time in some oncology and non-oncology trials.\cite{magirr2021non} Figure \ref{fig:KM} shows Kaplan-Meier curves for three oncology trials we examine in this paper as case studies, each exhibiting one of these three types of NPH patterns (delayed and diminishing effects, and crossing patterns); these studies and the rationale for examining them will be described in Section ``Case Studies''.

In NPH scenarios such as these, the Log-rank test loses efficiency, hence a wide range of alternative statistical methods for analyzing time-to-event data with NPH has been proposed in the literature. The weighted Log-rank test (WLRT), a generalized form of the Log-rank test, incorporates the Fleming and Harrington class of weights, \cite{fleming1981class} which uses two tuning parameters to assign different weights across time points and thus emphasizes particular segments of the survival curves.
The MaxCombo test is a method for combining multiple WLRTs that adaptively chooses the most suitable weights and provides more power than the traditional Log-rank test across various NPH scenarios.\cite{roychoudhury2021robust} The modestly weighted Log-rank test (MWLRT) is a new class of WLRT that controls the risk of concluding that a new treatment is more efficacious than the standard of care when it is in fact inferior.\cite{Magirr2019MWLRT} The weighted Kaplan-Meier test has been proposed for testing the equality of distributions in two-sample censored data and is more efficient than the Log-rank test in large samples.\cite{pepe1991weighted}

These non-parametric approaches provide useful alternative strategies to the Log-rank test in the presence of NPH, but they cannot provide interpretable estimates of the treatment effect. The Restricted Mean Survival Time (RMST) approach \cite{LuTian} defines RMST as the expected survival time up to a specific time point, corresponding to the area under the survival curve. Based on RMST, the treatment effect can be summarized as the difference or the ratio of RMSTs between two randomized arms and provides an alternative interpretation of the treatment effect as a difference or ratio of ``life expectancy'' up to that specific time point. 

However, all these methods have limitations. Global methods such as the Log-rank test and the RMST difference ($\Delta$RMST) test may fail to detect differences between two arms when positive and negative treatment effects at different times cancel each other out,\cite{davis2011caution} and misleading conclusions can be drawn by blindly accepting large p-values without examining the survival curves, considering their scientific and clinical significance, and assessing to what extent they violate a statistical test's assumptions. Recently, parametric models have been increasingly favored by researchers because of their interpretable parameters,  simplicity in studying relative times,\cite{cox2007ggm} flexibility in modeling time-varying effects, and non-reliance on the PH assumption. Additionally, they allow for a full characterization of the flexible hazard functions.

In this paper, we focus on testing the null hypothesis of identical survival curves against specific alternative hypotheses representing different NPH patterns (delayed and diminishing effects, and crossing pattern) by simulating data closely resembling survival curves from published IO trials. \cite{Shen2022nph} It is widely recognized that the Log-rank test may lead to severely underpowered studies in NPH situations. Using extensive simulations, we aim to identify good candidates for NPH resilience that exhibit acceptable power under common patterns of NPH while maintaining the type I error rate close to the nominal level.\cite{royston2020simulation} Additional simulation studies demonstrate the risk of using the MaxCombo test with crossing survival curves. Along with the MaxCombo test and the $\Delta$RMST test discussed above, our simulation studies also evaluate the Generalized Gamma model (GGM) \cite{cox2007ggm} and the Generalized F model (GFM).\cite{cox2008gfm} The Generalized Gamma (GG) family contains all four major forms of hazard functions (monotonically increasing and decreasing, concave up, and concave down) and the majority of the prevalent parametric survival distributions (e.g., exponential, Weibull, log-normal, and gamma), making it very helpful for estimating both individual hazard functions and relative hazards and times. The Generalized F (GF) family additionally includes the log-logistic distribution, providing added flexibility for parametric modeling.  In keeping with previous work in the field, we use power and type I error rates to compare the performance of these methods under various NPH scenarios seen in IO trials, and we also propose and implement novel metrics for assessing the time-dependent bias in treatment effect estimates to provide a more comprehensive evaluation of these methods in NPH scenarios. Recommendations under each NPH scenario are developed by examining the type I error rate, power, and time-dependent bias associated with each statistical approach.


The remainder of this article is structured as follows: In Section ``Methods'', we further describe all the tests being considered for PH assumption and treatment effects, and the metrics and parameters used to evaluate them. In Section ``Case Studies'', we report in detail the information and characteristics of the three published oncology trials used in this paper. In Section ``Simulation'', we explain our data-generating process when using underlying data from oncology trials and describe the setting and procedures used in our simulations of the three aforementioned NPH scenarios. Additionally, three scenarios representing the cancel-out effect (with the magnitudes of both the early and late effects increasing across scenarios) are introduced and described in detail. In Section ``Results'', we present the results of the simulations (including type I error, power, and time-dependent bias) from all the previously discussed models under the aforementioned scenarios. Section ``Discussion'' contains a discussion and summary of our findings along with some resulting recommendations. Conclusions are also presented in Section ``Discussion''.

\section{Methods}
The Grambsch-Therneau (G-T) test and Schoenfeld's global test are used to assess the non-proportionality of hazards in each of the three recently concluded oncology clinical trials. \cite{Shen2022nph} The Log-rank test is commonly used for the analysis of time-to-event trials and is therefore considered the benchmark for other methods in our comparisons. In this paper, we systematically evaluate the performance of several alternative statistical methods, including the MaxCombo test, the RMST difference ($\Delta$RMST) test, the Generalized Gamma model (GGM), and the Generalized F model (GFM) against the Log-rank test in both PH and NPH scenarios constructed based on three oncology trials. The statistical methods we evaluate are briefly described below.

\subsection{Tests of the Proportional Hazards Assumption}
The proportionality of hazards is the key assumption underlying both the Log-rank test and the Cox model when dealing with time-to-event data. Violation of this assumption can lead to biased estimates and incorrect inferences. Therefore, it is highly desirable to evaluate the proportionality of hazards before fitting the Cox model to survival data. 

Here we provide a brief overview of two types of tests for the evaluation of PH assumption based on Schoenfeld residuals.\cite{GTtest} The Schoenfeld residual is a measure of the difference between the observed and expected values of a covariate in a Cox model over time, where the expected value is calculated based on the distribution of the covariate in the population at risk. Given $d$ events occurring at distinct times $t_1,\dots,t_d$ (i.e., no ties present in the data), the Schoenfeld residual is defined as 
$$\boldsymbol r_s(\boldsymbol\beta)=\boldsymbol X_{s}-\Bar{\boldsymbol X}_s(\boldsymbol\beta),\ s=1,\dots,d,$$
where $\boldsymbol X_{s}$ is the observed covariate vector at time $t_s$ and $\Bar{\boldsymbol X}_s(\boldsymbol\beta)$ is the expected value (over the risk set just before time $t_s$) of the covariate vector at time $t_s$ given the constant vector of coefficients $\boldsymbol \beta$ estimated by the Cox model.

\subsubsection{Grambsch-Therneau (G-T) Test}
The Grambsch-Therneau (G-T) test\cite{GTtest,xue2020online} is based on the correlation between Schoenfeld residuals and some function of time. It takes the idea of the Schoenfeld residual one step further by allowing for the detection of time-varying covariate effects. It involves fitting a series of modified Cox regression models with time-varying coefficients and comparing the log-likelihood ratios of these models to the original model. If the likelihood ratio test is significant, this indicates that there is a time-varying effect of the covariate on the hazard ratio, violating the proportional hazards assumption.

\subsubsection{Schoenfeld's Global Test}

Schoenfeld's global test \cite{schoenfeld1980chi,abeysekera2009use} simultaneously evaluates the proportional hazards assumption for all covariates included in the Cox model. 
The test is based on the correlation between the Schoenfeld residuals and time for each covariate and a test statistic following the chi-square distribution with degrees of freedom equal to the number of covariates is constructed using the covariance matrix of the residuals. A significant correlation indicates at least one covariate violates the proportional hazards assumption. The test is particularly useful when there are multiple covariates in the model.

\subsection{Tests based on the Log-rank Test}



\subsubsection{The Log-rank Test}

The Log-rank test is the optimal (most powerful) test among all tests under PH. The power of the Log-rank test is reduced when the PH assumption is substantially violated. The general form of the Log-rank test statistic is defined as
$$Z(\omega) = \frac{U(\omega)}{\mathrm{se}\left[U(\omega)\right]},$$
where 
$$U(\omega) = \sum_t \omega\:(t) \left[\Delta N_1(t)-\frac{Y_1(t)}{Y(t)}\ \Delta N(t)\right] ,$$
$$\mathrm{se}\:(U(\omega)) = \sum_t \omega^2(t) \left[\frac{Y_1(t)\: Y_0(t)}{Y^2(t)}\right]\left[\frac{Y(t)-\Delta N(t)}{Y(t)-1}\right] \Delta N(t) ,$$
and $t$ indicates distinct event times, $\omega\:(t)$ is a weighting function across all event times (when $\omega\:(t) \equiv 1$, the above definition corresponds to the standard, unweighted Log-rank test), and $U(\omega)$ sums over the differences between the observed and expected number of events in the treatment group at all distinct event times. $Y_i(t)$ is the number of individuals at risk at time $t$ in group $i$ (in our examples, $i=1$ refers to the experimental treatment group and $i=0$ denotes the standard of care group), and $\Delta N_i(t)$ is the number of events observed within the $[t,t + \delta]$ time interval in group $i$. $Y(t)$ and $\Delta N(t)$, without subscripts, are the same metrics summed across both groups: $Y(t) = Y_0(t) + Y_1(t)$ and $\Delta N(t) = \Delta N_0(t) + \Delta N_1(t)$. The Log-rank statistic asymptotically follows a standard normal distribution under the null hypothesis.

The Fleming-Harrington $G\:(\rho,\gamma)$
family of weight functions \cite{fleming2011counting} can be used to adjust the weighting function of the general Log-rank test:
$$\omega\:(t)=\left[\hat S(t)\right]^{\rho}\left[1-\hat S(t)\right]^{\gamma},\ \rho \ge 0,\ \gamma \ge 0 ,$$
where $\hat S(t)$ is the Kaplan-Meier estimate of the survival function at time $t$  for the pooled survival data, and $(\rho,\gamma)$ enables the weighted test to focus on differences at various parts of the survival curves. We summarize commonly used weight functions in Table \ref{Gtable}. Note that when $(\rho,\gamma)=(0,0)$ and $\omega\:(t)=1$, we obtain the standard, unweighted Log-rank statistic.

\begin{table}[ht]
\small\sffamily\centering
\caption{Common choices of  $(\rho,\gamma)$ for the Fleming-Harrington $G\:(\rho,\gamma)$ family. \label{Gtable}}
\begin{tabular}{lllll}
\toprule
& & $(\rho, \gamma)$ & $\omega\:(t)$ & Type of test \\
\midrule
$\omega_0:$ && (0, 0) & 1 & Standard Log-rank test \\
$\omega_1:$ && (1, 0) & $\hat S(t)$ & Test of early difference \\ 
$\omega_2:$ && (0, 1) & $1-\hat S(t)$ & Test of late difference\\
$\omega_3:$ && (1, 1) & $\hat S(t)\left[1-\hat S(t)\right]$ & Test of middle difference\\
\bottomrule
\end{tabular}
\end{table}

\subsubsection{The MaxCombo Test}
The main objective of the MaxCombo test is to identify the maximum value of the test statistic in a set of weighted tests. The inclusion/exclusion of certain weighted tests often depends on the specific research question, and tests are selected on a case-by-case basis. Here, we consider the MaxCombo test utilizing the last three weighted Log-rank tests with the Fleming-Harrington family of weighting functions $\omega_1$, $\omega_2$, and $\omega_3$ described in Table \ref{Gtable}.\cite{fleming2011counting} The MaxCombo test statistic can be then described as 
$$Z=\max\left[Z(\omega_1),\: Z(\omega_2),\: Z(\omega_3)\right],$$
where $(\omega_1,\ \omega_2,\ \omega_3)$ are the weights corresponding to $(\rho,\gamma)=\left\{(0,1),\: (1,0),\:(1,1)\right\}$ (sensitive to delayed effects, diminishing effects, and middle effects, respectively), and $Z(\omega_k)$ (where $k \in \{1,2,3\}$ indexes the weights) is the standardized weighted Log-rank statistic defined above. The vector $\left(Z(\omega_1), Z(\omega_2), Z(\omega_3)\right)$ follows a multivariate normal distribution with a mean vector of 0 and an identity covariance matrix under the null hypothesis.  


\subsection{Test based on the Restricted Mean Survival Time Difference}

The restricted mean survival time $\mu$ is defined as the mean of the survival time $\min\:(T, t^*)$ limited to some right-censoring horizon $t^*>0$ \cite{andersen2004regression} where $T$ is the event time. Analytically, the value equals the area under the survival curve $S(t)$ evaluated from $t=0$ to $t=t^*$:
$$\mu=E\:\left(\min\:(T, t^*)\right)=\int_0^{t^*}S(t)\ dt.$$

In case $T$ is time to death, this quantity can be interpreted as a ``$t^*$-year life expectancy''. In a two-arm clinical trial with survival functions $S_0(t)$ and $S_1(t)$ of the control and treatment arms respectively, the difference in RMST ($\Delta$RMST) between arms, subject to right-censoring at the time $t^*$, is given by
$$\Delta RMST\:(t^*)=\int_0^{t^*}\left[S_1(t) -S_0(t)\right]\ dt.$$

The associated test statistic is then defined as 
$$Z(t^*)=\frac{\widehat{\Delta RMST\:(t^*)}}{\mathrm{se}\left(\widehat{\Delta RMST\:(t^*)}\right)}.$$

The use of the $\Delta$RMST as an alternative metric to the treatment effect, as well as the associated hypothesis test, have been demonstrated in the literature.\cite{Royston2011RMST,royston2013restricted} Using the ratio rather than the difference of the two integrated survival functions has also been explored in the literature.\cite{TrinquartRMSTratio} In this paper, the earlier of the maximum event times from each group is used as $t^*$ for the $\Delta$RMST test.\cite{Tianselecttimewindow} 

\subsection{Parametric Tests of Treatment Effects}
We evaluate the performance of flexible parametric survival models such as the Generalized Gamma model and the Generalized F model in various PH and NPH scenarios. 
The usefulness of the Cox model might be limited when we are also interested in estimating the acceleration factor, not hazard ratio, as the appropriate measure of association.\cite{cox2007ggm}

\subsubsection{The Generalized Gamma Model (GGM)}

The Generalized Gamma distribution $\mathrm{GG}(\beta,\sigma,\tau$) \cite{cox2007ggm} extends the two-parameter gamma distribution by introducing an additional parameter. Its probability density function \cite{cox2007ggm, jackson2010survival} is 
$$f_{GG}(t; \beta, \sigma, \tau) = \frac{|\tau|}{\sigma t \Gamma(\tau^{-2})} \left[ \tau^{-2} \left( e^{-\beta} t \right) ^{\tau/\sigma} \right]^{\tau^{-2}} \exp \left[ -\tau ^{-2} \left( e^{-\beta} t \right) ^{\tau/\sigma} \right]$$
where the location parameter $\beta$ scales time by $e^{-\beta}$, the scale parameter $\sigma$ sets the interquartile ratio for a fixed $\tau$ independently of $\beta$, and the parameter $\tau$ affects the shape of the distribution and can decide the type of the hazard function together with $\sigma$: when $\tau = \sigma$, this reduces to $\mathrm{Gamma}(k = \sigma^{-2}, \theta = \sigma^2 e^{\beta})$. The GGM also includes three widely used parametric survival models, Weibull ($\tau=1$), exponential ($\tau =\sigma = 1)$, and log-normal ($\tau=0$), as special cases and is often used to determine which parametric model is appropriate for a given data set. \cite{box2004event} It is a flexible model that accommodates all four common types of hazard function (i.e., monotonically increasing, monotonically decreasing, concave up, and concave down shapes). 


The median survival time (MST) can be computed from the quantile function for $\mathrm{GG}(\beta,\sigma,\tau)$: \cite{cox2007ggm}
$$\log[t_{GG(\beta,\sigma,\tau)}(0.5)]= \beta +\sigma \log[t_{GG(0,1,\tau)}(0.5)], $$
where $t_{GG(0,1,\tau)}(0.5)$ is $[\tau^2 \Gamma^{-1}(0.5;\tau^{-2})]^{1/\tau}$ and  $\Gamma^{-1}(0.5;\tau)$ is the quantile function of the gamma distribution with mean $\tau$.   Covariates can be modeled through $\beta = \beta_0 + \beta_1X_1 + \dots + \beta_nX_n$ to form a classical accelerated failure time (AFT) model \cite{cox2007ggm} with acceleration factors $\exp(-\beta_i)$ in terms of the covariates of interest. 
Test statistics can be obtained from a Wald test:
$${\displaystyle W={\frac {{({\widehat {\beta_i}}-\beta _{i0})}^{2}}{\operatorname {Var} ({\hat {\beta_i}})}}}.$$
This GGM based on the classical AFT form can be implemented using the R package ``flexsurv'', \cite{jackson2016flexsurv} which offers estimates for parameters such as $\beta$ (including coefficients for covariates), $\sigma$, and $\tau$. These estimates enable the precise estimation of survival or hazard functions for specific groups or individuals, contingent on their covariate values \cite{jackson2016flexsurv}.



\subsubsection{The Generalized F Model (GFM)}

The Generalized F distribution $\mathrm{GF}(\beta, \sigma, m_1, m_2)$  \cite{cox2008gfm} is a four-parameter family, generalizing the central F distribution with non-integer degrees of freedom $(2m_1, 2m_2)$ by adding location ($\beta$) and scale ($\sigma>0$) parameters according to the standard AFT model.\cite{wei1992accelerated} The probability density function is \cite{jackson2010survival}
$$f_{GF}(t;\beta, \sigma, m_1, m_2) = \frac{\delta(m_1/m_2)^{m_1} e^{m_1 w}}{\sigma t(1 + m_1 e^{w}/m_2)^{(m_1+m_2)} B(m_1, m_2)}$$
where $\delta = (m_1^{-1}+m_2^{-1})^{1/2}$, $w = (\log(t) - \beta)\delta/\sigma$, and the Beta function $ \mathrm {B}(m_{1},m_{2})=
\int _{0}^{1}t^{m_{1}-1}(1-t)^{m_{2}-1}\ dt$. To keep consistency with the GGM, a more stable parameterization $\mathrm{GF}\ (\beta, \sigma, q, p)$ \cite{prentice1975discrimination} replaces $0<m_1, m_2<\infty$ by alternative shape parameters 
$$q = (\frac{1}{m_1} - \frac{1}{m_2})(\frac{1}{m_1} + \frac{1}{m_2})^{-1/2},\ \ p=\frac{2}{m_1+m_2},$$
where $-\infty < q < \infty$ and $p<0$. Equivalently, 
$m_1 = 2(q^2 + 2p + q\delta)^{-1}$, and $m_2 = 2(q^2 + 2p - q\delta)^{-1}$.

The parameters $(\beta, \sigma, q, p)$ have similar interpretations to those in the GGM: $\beta$ is the time scalar, and $\sigma$, $q$, and $p$ decide the shape of hazard functions together. In the limiting case $p=0, q=\tau$ (i.e. the shape parameter of the GGM), $f_{GF}(t;\beta,\sigma,0,\tau) = f_{GG}(t;\beta,\sigma,\tau)$ and we recover the Generalized Gamma PDF above; more refined classifications can be further obtained by modifying $\tau$. The GFM also includes the generalized log-logistic ($q=0$) and log-logistic ($q=0, p=1$) parametric survival models as special cases. \cite{cox2008gfm} This allows the GFM to fit decreasing and concave down-shaped hazard functions better, particularly those that are decreasing but not monotone. 

As with GGM, the covariates can be modeled through $\beta$ and the test statistic is:
$${\displaystyle W={\frac {{({\widehat {\beta_i}}-\beta _{i0})}^{2}}{\operatorname {Var} ({\hat {\beta_i}})}}}.$$ 

The GFM can also be implemented using the R package ``flexsurv'', \cite{jackson2016flexsurv} which offers estimates for $\beta$ (including coefficients for covariates),  $\sigma$, $q$, and $p$, allowing for flexible interpretations.

 

\subsection{Time-Dependent Bias}

In order to provide a better evaluation of the overall performance of the above five methods in addition to the power and type I error rate, we calculated time-dependent bias, which was often ignored in previous papers \cite{ananthakrishnan2021critical, lin2020alternative}.  Here, we define the time-dependent bias of a quantity $\theta(t)$ (which could be the $\Delta$RMST over time, the survival probability difference over time, or the median survival time in both arms; these are all widely used to evaluate and compare the treatment effects) as 
$$\Delta \theta (t) = E\left[\hat \theta(t)\right]-\theta(t),$$ 
where $\theta(t)$ is the true value from the simulation and $E\left[\hat \theta(t)\right]$ is the expectation of the estimated value  $\hat\theta(t)$ from the Cox model, GGM, or  GFM.


\section{Case Studies}

The three case studies we consider  were selected to represent three different types of NPH patterns. More details on these studies can be found in the Supplementary materials. Individual patient-level data, including event times and censoring information, were reconstructed based on the published survival curves of these studies using the method described by Patricia Guyot et al. \cite{Guyot2012dataconstruct}

\begin{table}[th]
\small\sffamily\centering
\caption{Information summary of three oncology studies. The p-values of the G-T test and the sample size information come from Table A.1 in Yuan-Li Shen's paper.\cite{Shen2022nph} The p-values of Schoenfeld's global test were calculated using our reconstructed data. \label{studyinfo}}
\begin{tabular}{cccc}
\toprule
 & FIRST & INO-VATE & GOG-0218  \\
\midrule
Primary endpoint & Progression-free Survival  & Overall Survival & Progression-free Survival   \\
Sample size & 1082 & 326 & 1248 \\
NPH type & Early Crossing & Delayed Effect & Diminishing Effect  \\
 & (Delayed Effect) & & (Late Crossing) \\
G–T test & $p < 10^{-6}$ & $p$ = 0.50 & $p < 0.05$    \\
Schoenfeld's global test & $p < 10^{-6}$ & $p$ = 0.35 & $p < 0.05$  \\ 
\bottomrule
\end{tabular}
\end{table}

\paragraph{Early crossing: FIRST}

FIRST\cite{benboubker2014lenalidomide} (Frontline Investigation of Revlimid Plus Dexamethasone versus Standard Thalidomide) was a multicenter, randomized (1:1:1), open-label,
three-arm study comparing lenalidomide+low-dose dexamethasone for eighteen 28-day cycles (Rd continuous, $n$ = 535) or continued Rd beyond eighteen 28-day cycles (Rd18, $n$ = 541) versus melphalan+prednisone+thalidomide (MPT) control arm ($n$ = 547) in newly diagnosed multiple myeloma patients. The primary endpoint was progression-free survival (PFS).

\begin{figure}
\centering
\subfigure{
\includegraphics[width=2.5in]{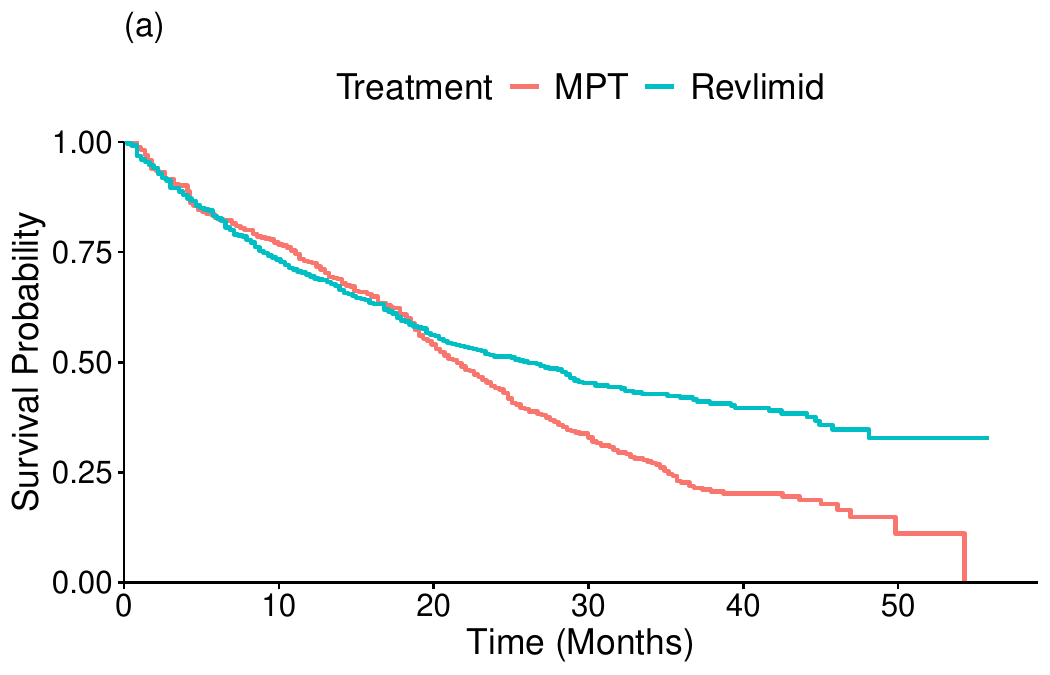}
    \label{study1-re}
}
\subfigure{
\includegraphics[width=2.5in]{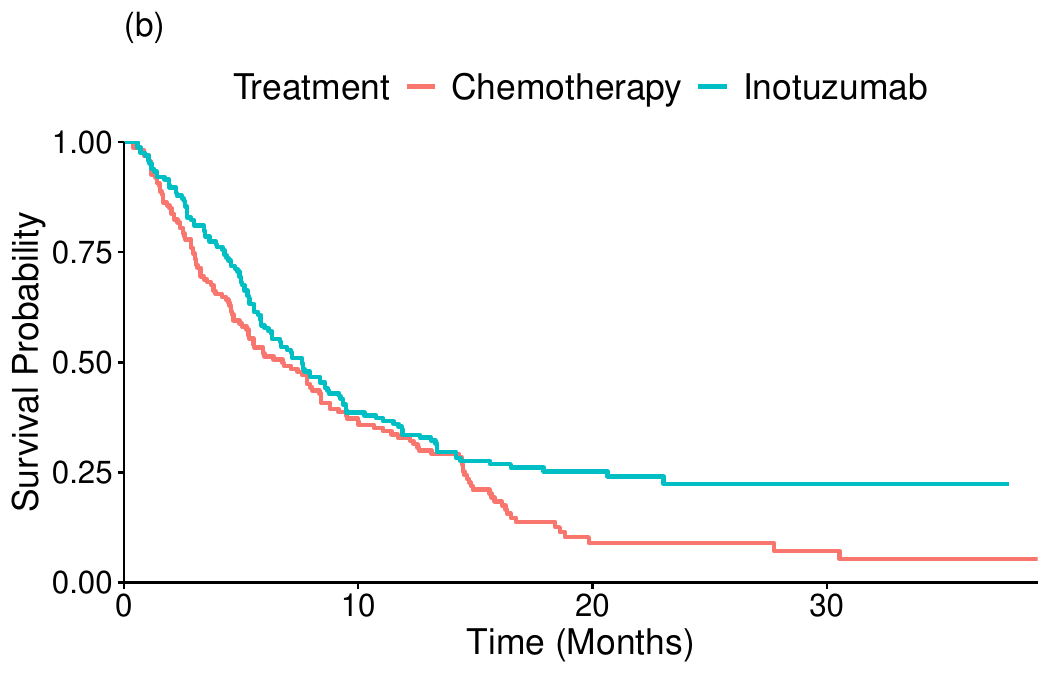}
    \label{study2-re}
}
\subfigure{
\includegraphics[width=2.5in]{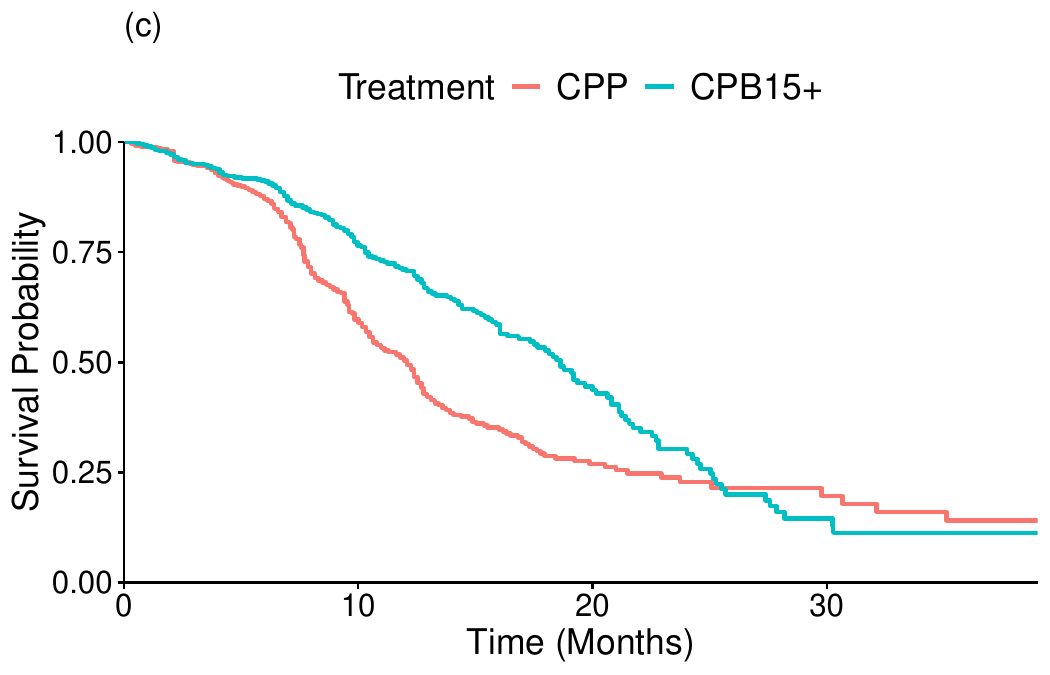}
    \label{study3-re}
}
\caption{Kaplan-Meier curves for three case studies using the reconstructed data. (a) Progression-free survival probability in FIRST; (b) Overall survival probability in INO-VATE; (c) Progression-free survival probability in GOG-0218.}
\label{fig:KM}
\end{figure}

Our comparison of interest was between the Rd continuous treatment arm and the MPT control arm.  In Figure \ref{study1-re}, we see the curves for PFS in the Rd continuous and MPT arms have a slight separation for the first roughly 18 months, but then the hazard ratio drastically changes, demonstrating an early crossing of the PFS curves. 

\paragraph{Delayed effect: INO-VATE}

INO-VATE\cite{kantarjian2016inotuzumab} (Inotuzumab Ozogamicin trial to investigate Tolerability and Efficacy) was a multicenter, randomized (1:1), open-label study comparing inotuzumab ozogamicin ($n$ = 164) to investigator’s choice of standard intensive chemotherapy ($n$ = 162) in adult patients with relapsed or refractory CD33+ acute lymphoblastic leukemia. The primary endpoint was overall survival (OS).

This case study represents the delayed effect scenario; the hazard ratio is close to 1 for the first roughly 13 months, then drastically increases in favor of the experimental treatment over the standard of care (see Figure \ref{study2-re}).  

\paragraph{Diminishing effect: GOG-0218}

GOG-0218 \cite{burger2011incorporation} \cite{GOG-0218info} was a multi-center, randomized (1:1:1), double-blinded, three-arm study comparing carboplatin and paclitaxel with concurrent bevacizumab, followed by bevacizumab single agent (CPB15+, $n$ = 623); carboplatin and paclitaxel with concurrent bevacizumab but no subsequent single-agent bevacizumab (CPB15, n = 625); and carboplatin and paclitaxel only as the control arm (CPP, n = 625). The goal of the study was to evaluate the effect of adding bevacizumab to carboplatin and paclitaxel for the treatment of patients with stage III or IV epithelial ovarian, fallopian tube, or primary peritoneal cancer following initial surgical resection. The primary endpoint was progression-free survival (PFS).

Focusing on the CPB15+ and the CPP arms, Figure \ref{study3-re} shows that the KM curves for PFS represent a diminishing effect scenario with a drastic increase in HR in favor of the control arm such that the two treatment curves cross roughly 26 months into the study.

\section{Simulation}
In this section, we explain our procedures for generating data that closely resemble the survival curves from the three published oncology trials, \cite{Shen2022nph}. We also describe the procedures for simulating three NPH scenarios based on the aforementioned trials and of three scenarios representing the cancel-out effect. The performance of these different methods was evaluated based on their statistical power, type I error rate, and time-dependent bias defined in Section ``Time-Dependent Bias''.

\subsection{Case Study Scenarios}
Data were simulated with piecewise exponential models to closely resemble the published survival curves from these three case studies. 
Within each case study, the intervals and knots were set to be the same for the treatment and the control arms. Piece-wise hazard rates $\lambda_{ij}$ (arm $i=0,1$; interval $j=1,\ldots,J)$ and hazard ratios $\mathrm{HR}_j = \lambda_{1j}/\lambda_{0j}$ were computed for the $j$-th interval in each case study based on the restructured data and were used to simulate trials under a variety of scenarios. The parameters defining the piecewise exponential models used for data simulation are provided in Table \ref{simu-para}. The follow-up times used are 60, 42, and 42 months, and the sample sizes are 1082, 326, and 1248 for three trials respectively. Our simulated survival curves are reasonably close to those provided in the original study papers, as shown in Figure \ref{fig:study_ori} in the Supplementary materials.

Under each of the three case study scenarios, we simulated 10,000 trials using the piecewise exponential model with the piecewise hazard rates shown in Table \ref{simu-para} to represent each study-specific NPH pattern. Power was estimated under the alternative hypothesis at $\alpha$ = 0.05 (2-sided), while the type I error rate was computed under the null hypothesis of identical survival curves. The survival data under the null were generated using an exponential model with a rate of 0.1. Time-dependent bias defined in Section ``Time-Dependent Bias'' in terms of survival probability and RMST difference were computed for each simulated trial, at thirty time points equally spaced between the start and end of pre-specified follow-up. Median survival time for each simulated trial was estimated by searching for the root of 
the quantile function at 50\% quantiles for the Cox model, the GGM, and the GFM, respectively.

\begin{table}[ht]
\scriptsize\sf\centering
\caption{Simulation parameters for each study. $\lambda_0$ and $\lambda_1$ are the rates used in the piecewise exponential distribution in each arm, and intervals were set to be the same for the two arms. $\mathrm{HR} =\lambda_1/\lambda_0$ within each interval. \label{simu-para}}
\begin{tabular}{ccccccccccccc}
\toprule
Study & \multicolumn{4}{c}{FIRST (Early Crossing)} & \multicolumn{4}{c}{INO-VATE (Delayed Effect)} & \multicolumn{4}{c}{GOG-0218 (Diminishing Effect)} \\
\midrule
Parameters & $\lambda_0$ & $\lambda_1$ & Interval & HR & $\lambda_0$ & $\lambda_1$ & Interval & HR & $\lambda_0$ & $\lambda_1$ & Interval & HR \\
\cmidrule(lr){2-5} \cmidrule(lr){6-9} \cmidrule(lr){10-13}
 & 0.028 & 0.031 & [0,8) & 1.130 & 0.106 & 0.068 & [0,4) & 0.645 & 0.023 & 0.015 & [0,6) & 0.661 \\
 & 0.033 & 0.027 & [8,20) & 0.826 & 0.100 & 0.122 & [4,8) & 1.250 & 0.097 & 0.044 & [6,15) & 0.451 \\
 & 0.050 & 0.022 & [20,30) & 0.435 & 0.075 & 0.083 & [8,12) & 1.109 & 0.061 & 0.065 & [15,20) & 1.076 \\
 & 0.015 & 0.009 & [30,60] & 0.593 & 0.144 & 0.040 & [12,16) & 0.278 & 0.032 & 0.150 & [20,30) & 4.762 \\
 &       &       &        &       & 0.144 & 0.020 & [16,42] & 0.143 & 0.017 & 0.055 & [30,42] & 3.275 \\
\bottomrule
\end{tabular}
\end{table}

\subsection{Cancel-Out Effect Scenarios}

Studies have shown that crossing survival curves can be detrimental to the analysis of survival data using the classical Log-rank test and Cox proportional hazards method.\cite{davis2011caution} In order to understand how the cancel-out effect of crossing survival curves affects the performance of the aforementioned five analysis methods in Sections ``Tests based on the Log-rank Test'', ``Test based on the Restricted Mean Survival Time Difference'' and ``Parametric Tests of Treatment Effects'', we constructed three simulation scenarios with hazard ratios and knots designed specifically to make the early effect of one treatment and the late effect of the competing treatment ``cancel out.'' The parameters for each of the three scenarios are listed in Table \ref{cancelpara}. An exponential model with a low event rate of 0.1 per month was used to simulate data in the control arm, while a piecewise exponential model with knots at $t_1$ and $t_2$ was used for the treatment arm for its flexibility and efficiency in modeling time-varying treatment effects, making HR($t$) a time-dependent step function. The censoring rate was set as 0.01 from an exponential model. To achieve the desired cancel-out effect, we fixed a constant hazard ratio of 0.1 for the middle interval and gradually increased the hazard ratios in the first and the last interval to make the treatment effects increasingly pronounced. Knots were moved as needed to maintain overall balance. The trial sample size was set to 500 for each arm, and the total duration of each study was set to 24 months. 

\begin{figure}[!ht]
    \centering
    \includegraphics[width=5.5in]{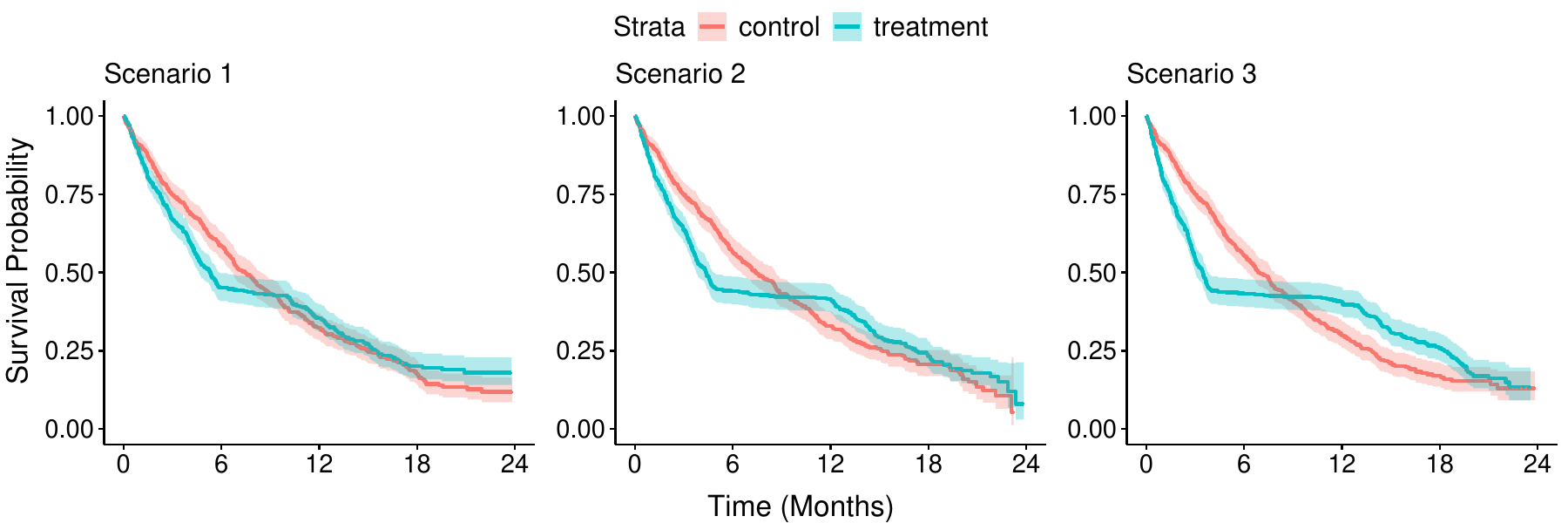}
    \caption{Simulated Kaplan-Meier curves of three cancel-out effect scenarios (larger effect from scenario 1 to 3, left to right). The hazard rate of the control group is 0.1 per month for each scenario while the parameters of the treatment group change as indicated in Table \ref{cancelpara}. The censoring rate is 0.01. The study duration for all scenarios is 24 months and the sample size is 500 for each arm.}
    \label{gradient}
\end{figure}

Kaplan-Meier curves for each scenario are shown in Figure \ref{gradient}; note the gradual increase in effect sizes from left to right.

\begin{table}[ht]
\small\sffamily\centering
\caption{Parameters of the cancel-out effect simulations used in the piecewise exponential distribution. From scenarios 1 to 3, the treatment effect increases. HR represents the hazard ratios within each interval, and the overall balance was maintained by adjusting knots (intervals) as needed. \label{cancelpara}}
\begin{tabular}{ccccccc}
\toprule
Scenario & 1 &  & 2 &  & 3 &  \\
\midrule
Parameters & Interval & HR  &  Interval  & HR & Interval & HR \\
\cmidrule(lr){2-3} \cmidrule(lr){4-5} \cmidrule(lr){6-7}
 & $[0,6)$ & 1.3 & $[0,5)$ & 1.6 & $[0,4)$ & 2.0 \\
 & $[6,10)$ & 0.1 & $[5,12)$ & 0.1 & $[4,13)$ & 0.1 \\
 & $[10,24]$ & 1.1 & $[12,24]$ & 1.2 & $[13,24]$ & 1.3 \\
\bottomrule
\end{tabular}
\end{table}

For each crossing pattern, we compared the power of the Log-rank test, the MaxCombo test, the $\Delta$RMST test, the GGM, and the GFM to evaluate their ability to detect time-specific and overall treatment effects. Based on the power and performance of our set of analysis methods in the presence of a cancel-out effect, we will make recommendations regarding which tests are reasonable choices when faced with crossing survival curves producing a small global effect.

\section{Results} 

\subsection{Results for the Three Simulated Case Studies} 
 First, we examined the PH assumption by Schoenfeld's global test for each case study using our reconstructed data.  G-T tests of these studies have been previously published,\cite{Shen2022nph} and the results of both sets of tests are shown in Table \ref{studyinfo}. For FIRST, both the G-T test ($p < 10^{-6}$) and Schoenfeld's global test ($p < 10^{-6}$) indicate significant violations of the proportional hazards assumption. For INO-VATE, although the G-T ($p=0.50$) and Schoenfeld's global tests ($p=0.35$) are not significant, non-proportional hazards are apparent upon visual inspection of the Kaplan-Meier curves in Figure \ref{fig:study_ori}. For GOG-0218, the non-proportional hazard pattern is both clear on inspection and confirmed by both NPH tests ($p<0.05$). 
 
\begin{figure}[ht]
    \centering
    \includegraphics[width = 5.5in]{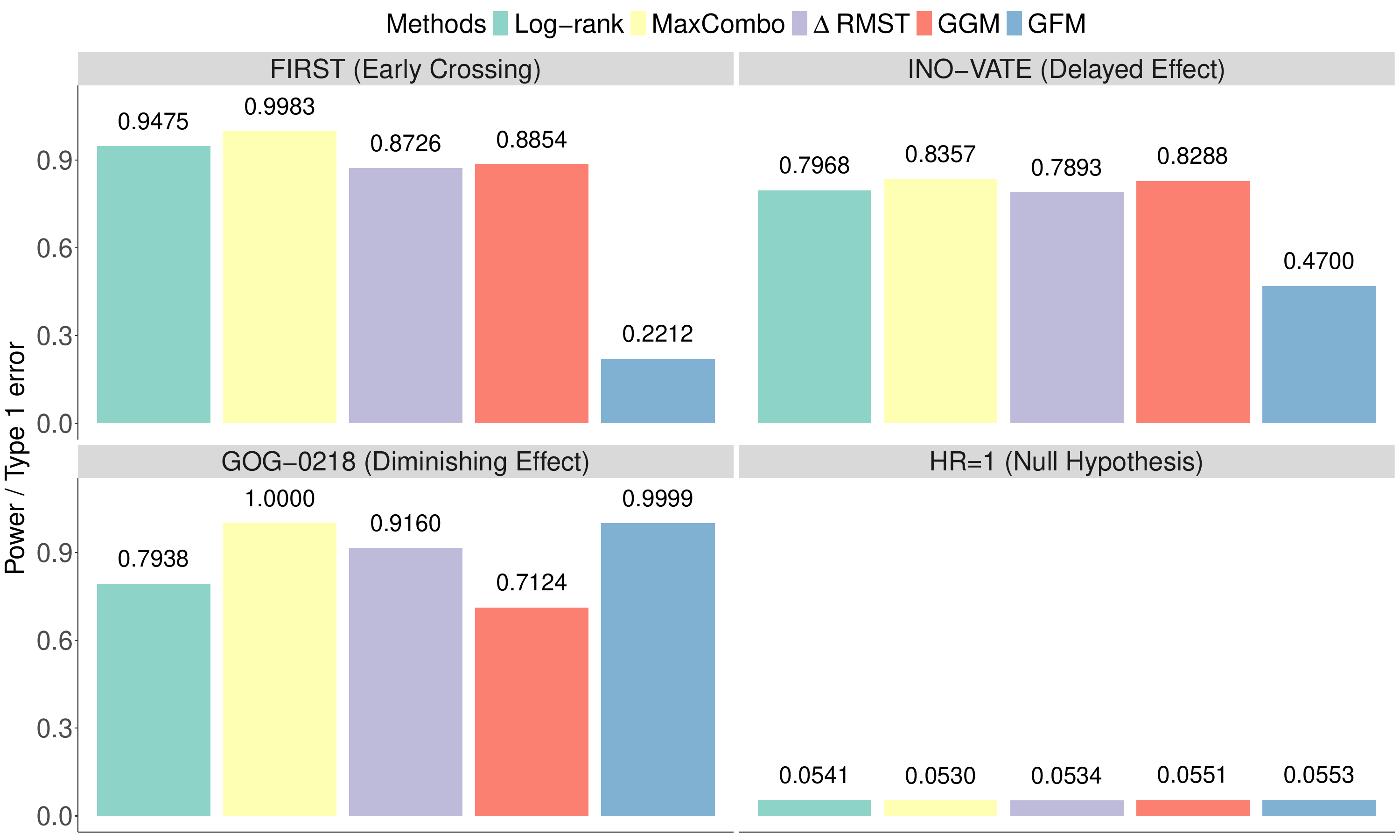}
    \caption{Power under three case studies with distinct NPH patterns and type I error under the null hypothesis, calculated from 10,000 replications.}
    \label{fig:power_type1error}
\end{figure}

We then estimated the empirical type I error rates of the five analysis methods under consideration (the Log-rank test, the MaxCombo test, the $\Delta$RMST test, the GGM, and the GFM) based on 10,000 trials simulated under the null hypothesis, using a two-sided significance level of 0.05.  As shown in Figure \ref{fig:power_type1error}, the performance of all five methods is reasonably close to the nominal, two-sided 5\% type I error rate. 

We present the estimated power from each case study (again based on 10,000 trials) for each analysis method in Figure \ref{fig:power_type1error}. For FIRST (early crossing scenario), the MaxCombo test has the highest power, close to one, followed by the Log-rank test with a power of 0.95. The $\Delta$RMST test and the GGM provide a lower and similar power of about 0.88, while the GFM is the worst at only 0.22, showing that the GFM has little ability to detect a group difference having this pattern. For INO-VATE (delayed effect scenario), the four methods other than the GFM all have power around 0.8, with the MaxCombo test and the GGM having slightly higher power than the other two. The GFM still has the worst performance with 0.47 power. For GOG-0218 (diminishing effect scenario), the MaxCombo test and the GFM perform the best, with power very close to 1.  The GFM's much better performance in this case seems in keeping with its capability at modeling decreasing but non-monotone hazard functions. The Log-rank and $\Delta$RMST tests are both affected by the survival curves' moderate crossing and have lower power, with less impact on the $\Delta$RMST test.  The GGM has the lowest power, 0.71, attesting to its relative disadvantage in detecting group differences in the diminishing effect pattern. 

As mentioned in Section ``Parametric Tests of Treatment Effects'', the GGM and the GFM are both flexible parametric models with the capability to accommodate different shapes of hazard functions, making them a good alternative to the Cox model in NPH scenarios. Table \ref{AFandHR} shows the regression coefficients, acceleration factors (AF), and hazard ratios (HR) derived from the Cox model, the GGM, and the GFM across the three oncology trials. Both the GGM and the GFM have AFT interpretations; consider the FIRST trial as an example. The hazard ratio is $\exp(-0.307)= 0.736$
based on the Cox model, suggesting the relative risk of death for patients in the treatment group compared to the control group is 0.736. Based on the AFT models, the acceleration factor for the treatment group vs control group is $\exp(-0.292)=0.747$, suggesting the median survival time for patients in the treatment group is about $x$ times that of patients in the control group. We observed that for all three trials, the GGM and GFM models are in close agreement in terms of the estimated acceleration factor, and all the models (AFT or Cox) suggest the treatment improves the survival experience of patients. 


\begin{table}[ht]
    \small\sffamily\centering 
    \caption{Summary table of regression coefficients, acceleration factors (AF), and hazard ratios (HR) from the Cox, the GGM, and the GFM across three clinical trials. AF is $\exp(-\beta)$ and HR is $\exp(\beta)$ because of the different reference groups.}
    \begin{tabular}{cccccccccc}
    \toprule
    Trial & \multicolumn{3}{c}{FIRST (Early Crossing)} & \multicolumn{3}{c}{INO-VATE (Delayed Effect)} & \multicolumn{3}{c}{GOG-0218 (Diminishing Effect)} \\
    \midrule
    & $\beta$ & AF & HR & $\beta$ & AF & HR & $\beta$ & AF & HR \\
    \cmidrule(lr){2-4} \cmidrule(lr){5-7} \cmidrule(lr){8-10}
    Cox & -0.307 & & 0.736 & -0.283 & & 0.754 & -0.456 & & 0.634 \\
    GGM & 0.292 & 0.747 & & 0.260 & 0.711 & & 0.293 & 0.746 & \\
    GFM & 0.292 & 0.747 & & 0.257 & 0.773 & & 0.397 & 0.672 & \\
    \bottomrule
    \end{tabular}
    \label{AFandHR}
\end{table}

\begin{figure}[ht]
    \centering
    \includegraphics[width=5.5in]{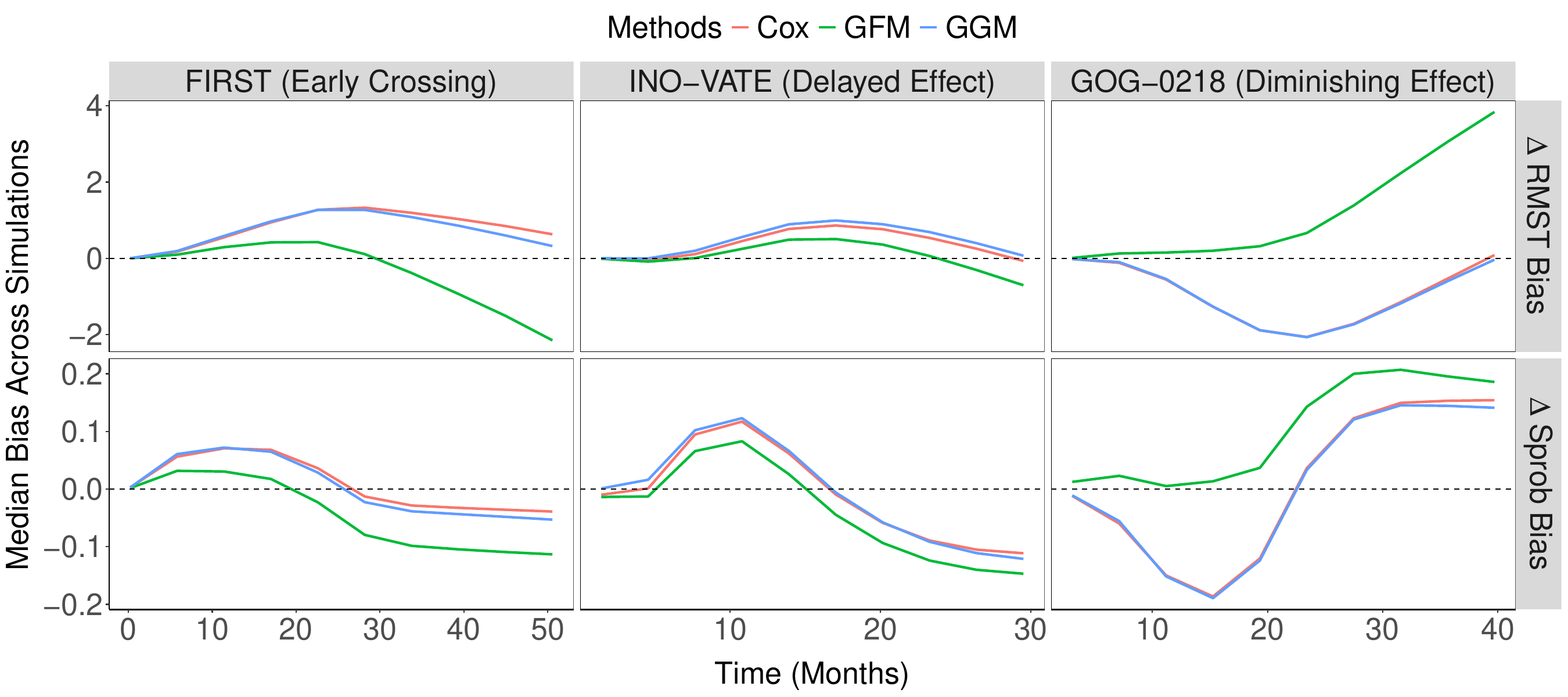}
    \caption{Median bias of RMST difference and survival probability difference (RMST, Sprob for short) for each model (Cox, GGM, and GFM) over time in each case study. Median bias is determined by the 10,000 replications of time-dependent bias at each time point, which is defined as $E\left[\hat\theta(t)-\theta(t)\right]$. Thirty equally spaced time points were selected between the start and end of the follow-up for each line plot. The black dotted lines in the median bias plots indicate a bias of 0.}
    \label{fig:allbias}
\end{figure}

\begin{figure}
    \centering
    \includegraphics[width=5.5in]{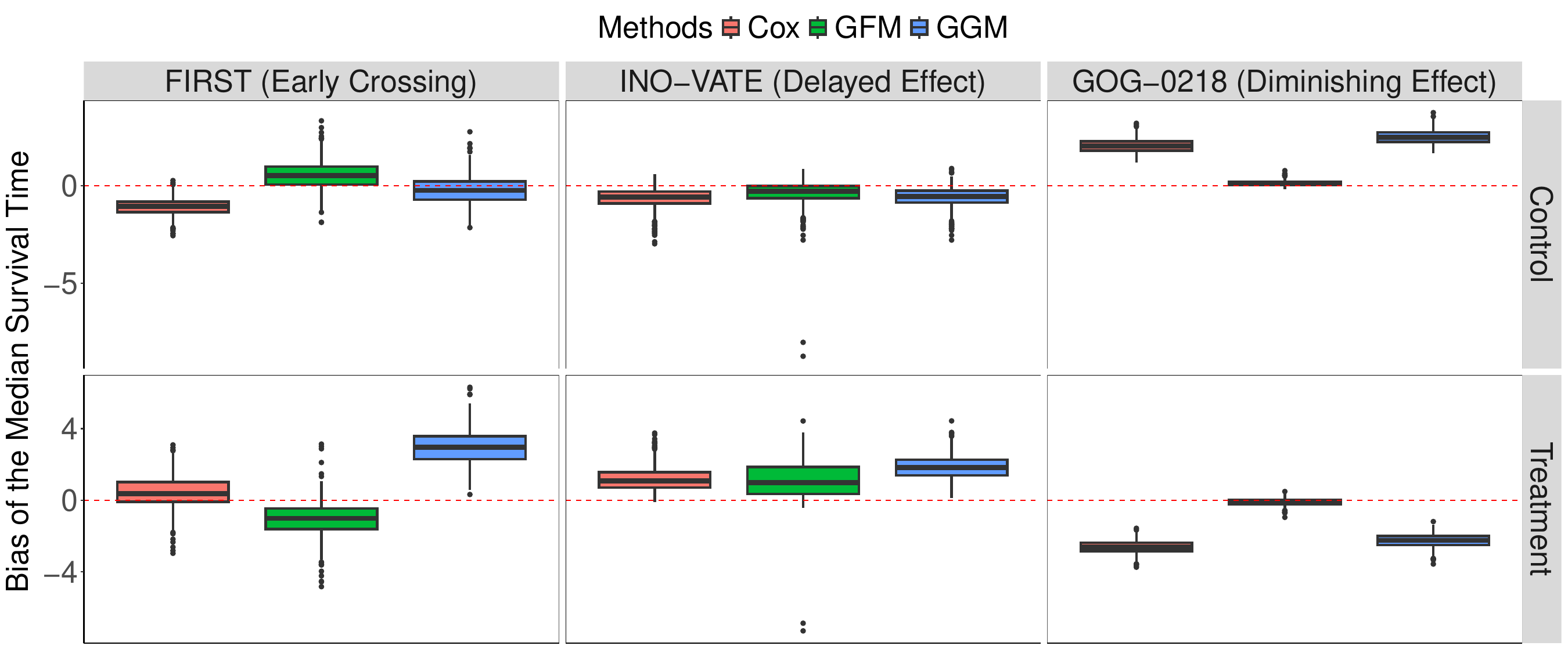}
    \caption{Bias of the median survival time in the control group and treatment group in each case study for the simple Cox model, GGM, and GFM. The red dotted lines indicate a bias of 0.}
    \label{fig:mst_bias}
\end{figure}

To provide a good overall description of the capabilities of these methods in estimating the true treatment effect, we also evaluated the time-dependent biases in the estimates produced by the parametric and semi-parametric approaches that provide interpretable treatment effect estimates. As introduced in Section ``Time-Dependent Bias'', time-dependent bias is computed from the Cox model, the GGM, or the GFM. The median biases in RMST difference and in survival probability difference over time in each case study are presented in Figure \ref{fig:allbias}, and Figure \ref{fig:mst_bias} shows the biases in the median survival time of two arms.

The RMST difference provides the average effect over a time period, while the survival probability difference focuses on the instantaneous effect at a specific point in time. The RMST difference can be seen as the integral of the survival probability difference. In Figure \ref{fig:allbias}, we see the estimates of the Cox model are very similar to those of the GGM model in all three cases, with differences only becoming apparent at the longest cutoff times. The median bias of the GFM is relatively small in the short term in each case, but in the longer term, it tends to produce a larger bias, indicating that the GFM does not fit the tail of the curve very well. In light of this pattern and the power results in Figure \ref{fig:power_type1error}, the GFM seems better suited to cases without a delayed effect.

In oncology research, the median survival time may be used to evaluate the efficacy of a new treatment by comparing it to a standard treatment or placebo. A longer median survival time indicates that the new treatment is more effective in prolonging survival. In Figure \ref{fig:mst_bias}, we see that the GFM's lesser bias in the early and mid-term and overall flexibility allow it to perform extremely well in both arms under each scenario, especially in GOG-0218, where it has a minimal bias with little variance. The performances of the GGM and the Cox model are somewhat worse, especially in GOG-0218. When the time until a particular event occurs is of primary interest, the GFM seems to be a good tool for estimating the median survival time accurately. 

Before moving on, we emphasize that these three case studies do not comprehensively represent all possible types of NPH, so while we hope that keeping these findings in mind can be of use, it remains important to exercise caution and consider the underlying research questions when arriving at a final conclusion.

\subsection{Results for the Simulated Cancel-Out Effects}

\begin{figure}[!ht]
    \centering
    \includegraphics[width=5.5in]{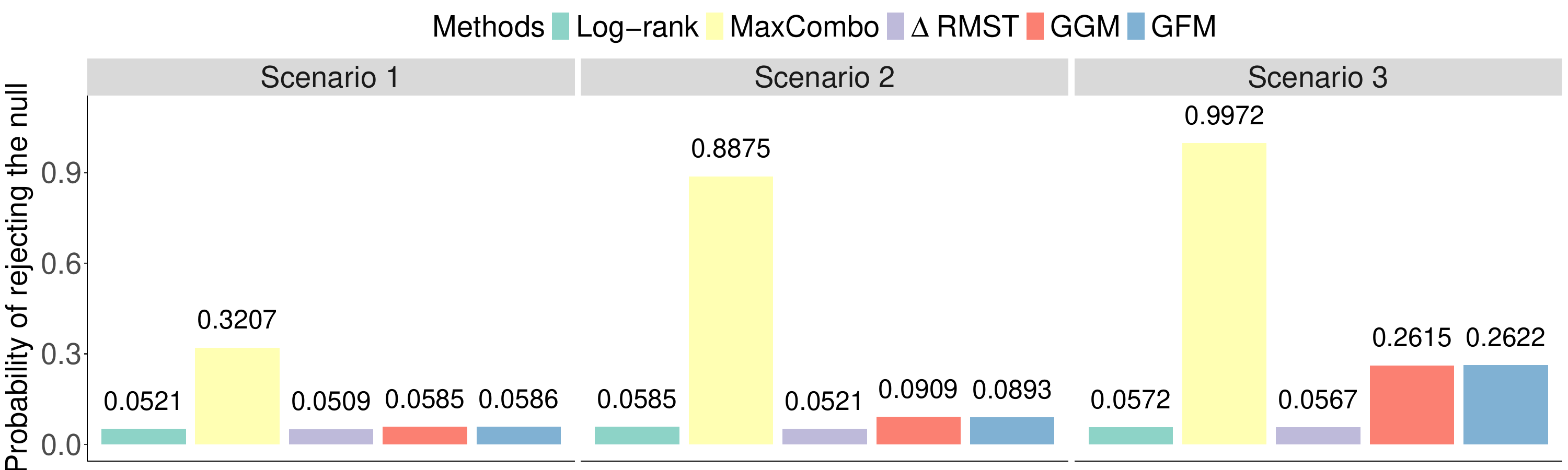}
    \caption{The power of different methods under the crossing hazards scenarios (scenarios 1-3 represent increasing crossing effect).}
    \label{gradientpower}
\end{figure}

In this section, we assess the performance of different analysis approaches by constructing three cancel-out effect scenarios. As expected, the crossing hazards in our simulations caused the net or global effects to be small,  rendering a verdict of no treatment effect that does not truly represent the data. Figure \ref{gradientpower} shows that the power of the MaxCombo test quickly approaches 1 as the differences in survival probability before and after the crossing increase, while the power of the Log-rank and $\Delta$RMST tests are consistently close to the nominal level ($\sim$.05) one would expect under the null. Such results indicate that all of these methods need to be used with caution. When treatment effects in opposite directions can completely cancel each other out, the global Log-rank and $\Delta$RMST tests are less likely to detect any difference between groups, followed by the two parametric models. This is not supprising as the net effect of treatment on survival in these crossing hazard scenarios are 0 and all these methods except max-Combo provide tests for the overall treatment effect. In comparison, the MaxCombo test amplifies piecewise differences and gives extremely high power as long as a local difference exists somewhere. 

Table \ref{tab:summ} summarizes all the methods and scenarios we evaluated in this paper.

\begin{table}[ht]
    \centering
    \scriptsize\sffamily 
    \caption{Summary table for five methods evaluated in this paper. Column "Model" refers to the parametric or nonparametric method. The check mark indicates this method can be a reasonable choice under the corresponding scenario while the question mark reminds us that we should be cautious when using.\label{tab:summ}}
    \begin{tabular}{cc>{\centering\arraybackslash}p{1.5cm}>{\centering\arraybackslash}p{1.5cm}>{\centering\arraybackslash}p{1.5cm}>{\centering\arraybackslash}p{3.5cm}}
    \toprule
    Method  & Model & Early Crossing & Delayed Effect & Diminishing Effect & Crossing Survival Curves with \newline No Global Treatment Effect \\
    \midrule
    Log-rank & Nonparametric & $\checkmark$ & $\checkmark$ & $\checkmark$ & \\
    MaxCombo & Nonparametric & $\checkmark$ & $\checkmark$ & $\checkmark$ & $?$ \\
    $\Delta$RMST & & $\checkmark$ & $\checkmark$ & $\checkmark$ & \\
    Generalized Gamma & Parametric (AFT) & $\checkmark$ & $\checkmark$ & & \\
    Generalized F & Parametric (AFT) & & & $\checkmark$ & \\
    \bottomrule
    \end{tabular}
\end{table}

\newpage
\section{Discussion} 
Compared to the non-parametric and semi-parametric tests we considered, the GGM and GFM methods are not reliant upon the PH assumption, and a complete description of the hazard function can be obtained. \cite{cox2007ggm} In our simulations, the parametric models perform comparably to the others in general and perform better under some specific scenarios in terms of power, type I error, and bias. Both methods provide good control of type I error rates, and each has an effect pattern it excels at fitting. The GGM performs well in delayed effect scenarios, with power comparable to or greater than that provided by other methods and with a relatively smaller bias than the GFM. In contrast, the GFM is more suitable when the focus is on diminishing effects since its available hazard functions excel in fitting hazards that are overall decreasing but not monotone; with these effects, the fluctuations in the GFM's bias are more balanced around 0 in the early and middle stages. 


During our literature review of non-proportional hazard patterns in oncology trials, it became obvious that the MaxCombo test was the most often recommended analytical tool for its relative robustness across patterns in the absence of any prior knowledge regarding the shape of the pattern.\cite{Ray2020} If we focus solely on type I error and magnitude of power, the MaxCombo test indeed provides the best performance: In our simulations, the MaxCombo test was able to maintain type I error around the target level and was also the most powerful across all the investigated NPH scenarios (likely a consequence of its adaptive selection of the weights). However, a major conclusion from our simulations is that one must be very careful when using the MaxCombo test, as it can result in inflated power and unreliable conclusions. In particular, when confronted with crossing survival curves, the MaxCombo test will always detect a significant difference even if at the end of the study there is no detectable difference between the two investigated treatments; with this pattern, the MaxCombo test over-emphasizes any detectable differences between the groups that may occur at any time over the course of the study. This highlights the importance of pre-specifying the primary objective for the study and the specific hypothesis we want to test. Our artificial ``cancel-out effect'' scenarios show that it is important to specify a priori whether we want to be sensitive to any differences over time or focus on early effects (e.g., for some surgical interventions), late effects (e.g., for anti-viral therapy), middle effects (e.g., the high-dose dexamethasone for multiple myeloma \cite{rajkumar2010lenalidomide} and chemotherapy for advanced non–small-cell lung cancer \cite{nsclc2008chemotherapy}) or the global effect. In any case, when using the MaxCombo test, the set of weights has to be pre-specified during the design stage, based on prior knowledge and clinical relevance.

The Log-rank test is underpowered when the proportionality assumption is severely violated and the main objective is to detect local differences between the treatments. In the presence of crossing survival curves, both the Log-rank test and the $\Delta$RMST test will be affected by the change in the effect's direction and lose power to detect even a major group difference when those differences occur partway through follow-up but have partially or fully canceled out by the end of the study. Our most extreme demonstration of this was in our ``cancel-out'' scenarios, where the power of these tests dropped to the level of type I error, but the issue was also evident in the FIRST trial (early crossing) and in GOG-0218 (late crossing). Thus, these tests are best used when a study's focus is on global effects.  In INO-VATE, where the survival curves did not cross and the hazards had a milder violation of proportionality, the performance of the Log-rank and $\Delta$RMST tests came very close to the performance of the MaxCombo test.  This conjecture is consistent with the simulation results of Lin et al.\cite{lin2020alternative} on the performance of the MaxCombo test under various scenarios.

To summarize, when the survival curves cross, we should be cautious of using the MaxCombo test. The MaxCombo test is overly sensitive to local effects and can produce false positive results when the objective is really to detect an overall difference.  The GGM and the GFM can be good alternatives to non- and semi-parametric methods under both PH and certain NPH scenarios.


One potential limitation of this study is whether our findings generalize beyond the specific NPH types investigated in this analysis (delayed effects and crossing survival curves). To ascertain the robustness and applicability of the proposed methods, it would be necessary to conduct additional simulation studies incorporating a broader range of NPH patterns, as well as to apply the developed techniques to a wider range of real-world immuno-oncology randomized trial datasets. Additionally, we note that no single summary statistic fully captures treatment effects in non-proportional hazards, with the MaxCombo test offering limited interpretability.  Further investigation of how best to capture treatment effects and interpret the results of studies exhibiting NPH is beyond the scope of the present paper and the direction of our ongoing research. 

\begin{acks}
This work was supported by CTSA Grant Number UL1 TR0001863 from the National Center for Advancing Translational Science (NCATS), a component of the National Institutes of Health (NIH).
\end{acks}

\begin{dci}
    The authors declared no potential conflicts of interest with respect to the research, authorship, and/or publication of this article.
\end{dci}

\begin{sm}
Simulation codes for this article is available online: \href{https://github.com/XinyuIvy/NPH.git}{Github}.
\end{sm}

\bibliographystyle{SageV}
\bibliography{Bibliography.bib}

\begin{thebibliography}{10}
\providecommand{\url}[1]{\texttt{#1}}
\providecommand{\urlprefix}{URL }
\expandafter\ifx\csname urlstyle\endcsname\relax
  \providecommand{\doi}[1]{DOI:\discretionary{}{}{}#1}\else
  \providecommand{\doi}{DOI:\discretionary{}{}{}\begingroup \urlstyle{rm}\Url}\fi
\providecommand{\eprint}[2][]{\url{#2}}

\bibitem{schoenfeld1981asymptotic}
Schoenfeld D.
\newblock The asymptotic properties of nonparametric tests for comparing survival distributions.
\newblock \emph{Biometrika} 1981; 68(1): 316--319.

\bibitem{HodiIO}
Hodi FS, O'Day SJ, McDermott DF et~al.
\newblock Improved survival with ipilimumab in patients with metastatic melanoma.
\newblock \emph{New England Journal of Medicine} 2010; 363(8): 711--723.
\newblock \doi{10.1056/NEJMoa1003466}.
\newblock \urlprefix\url{https://doi.org/10.1056/NEJMoa1003466}.
\newblock PMID: 20525992, \eprint{https://doi.org/10.1056/NEJMoa1003466}.

\bibitem{wolchok2010ipilimumab}
Wolchok JD, Neyns B, Linette G et~al.
\newblock Ipilimumab monotherapy in patients with pretreated advanced melanoma: a randomised, double-blind, multicentre, phase 2, dose-ranging study.
\newblock \emph{The lancet oncology} 2010; 11(2): 155--164.

\bibitem{small2006placebo}
Small EJ, Schellhammer PF, Higano CS et~al.
\newblock Placebo-controlled phase iii trial of immunologic therapy with sipuleucel-t (apc8015) in patients with metastatic, asymptomatic hormone refractory prostate cancer.
\newblock \emph{J clin Oncol} 2006; 24(19): 3089--94.

\bibitem{alexander2018hazards}
Alexander BM, Schoenfeld JD and Trippa L.
\newblock Hazards of hazard ratios-deviations from model assumptions in immunotherapy.
\newblock \emph{The New England journal of medicine} 2018; 378(12): 1158--1159.

\bibitem{rahman2019divining}
Rahman R, Ventz S, Fell G et~al.
\newblock Divining responder populations from survival data.
\newblock \emph{Annals of Oncology} 2019; 30(6): 1005--1013.

\bibitem{disis2014mechanism}
Disis ML.
\newblock Mechanism of action of immunotherapy.
\newblock In \emph{Seminars in oncology}, volume~41. Elsevier, pp. S3--S13.

\bibitem{mok2009gefitinib}
Mok TS, Wu YL, Thongprasert S et~al.
\newblock Gefitinib or carboplatin--paclitaxel in pulmonary adenocarcinoma.
\newblock \emph{New England Journal of Medicine} 2009; 361(10): 947--957.

\bibitem{ananthakrishnan2021critical}
Ananthakrishnan R, Green S, Previtali A et~al.
\newblock Critical review of oncology clinical trial design under non-proportional hazards.
\newblock \emph{Critical Reviews in Oncology/Hematology} 2021; 162: 103350.

\bibitem{magirr2021non}
Magirr D.
\newblock Non-proportional hazards in immuno-oncology: Is an old perspective needed?
\newblock \emph{Pharmaceutical Statistics} 2021; 20(3): 512--527.

\bibitem{fleming1981class}
Fleming TR and Harrington DP.
\newblock A class of hypothesis tests for one and two sample censored survival data.
\newblock \emph{Communications in Statistics-Theory and Methods} 1981; 10(8): 763--794.

\bibitem{roychoudhury2021robust}
Roychoudhury S, Anderson KM, Ye J et~al.
\newblock Robust design and analysis of clinical trials with nonproportional hazards: a straw man guidance from a cross-pharma working group.
\newblock \emph{Statistics in Biopharmaceutical Research} 2021; : 1--15.

\bibitem{Magirr2019MWLRT}
Magirr D and Burman CF.
\newblock Modestly weighted logrank tests.
\newblock \emph{Statistics in Medicine} 2019; 38(20): 3782--3790.
\newblock \doi{https://doi.org/10.1002/sim.8186}.
\newblock \urlprefix\url{https://onlinelibrary.wiley.com/doi/abs/10.1002/sim.8186}.
\newblock \eprint{https://onlinelibrary.wiley.com/doi/pdf/10.1002/sim.8186}.

\bibitem{pepe1991weighted}
Pepe MS and Fleming TR.
\newblock Weighted kaplan-meier statistics: Large sample and optimality considerations.
\newblock \emph{Journal of the Royal Statistical Society: Series B (Methodological)} 1991; 53(2): 341--352.

\bibitem{LuTian}
Zhao L, Claggett B, Tian L et~al.
\newblock On the restricted mean survival time curve in survival analysis.
\newblock \emph{Biometrics} 2016; 72(1): 215--221.
\newblock \doi{https://doi.org/10.1111/biom.12384}.
\newblock \urlprefix\url{https://onlinelibrary.wiley.com/doi/abs/10.1111/biom.12384}.
\newblock \eprint{https://onlinelibrary.wiley.com/doi/pdf/10.1111/biom.12384}.

\bibitem{davis2011caution}
Davis M and Xie SX.
\newblock Caution: hazards crossing! using the renyi test statistic in survival analysis.
\newblock \emph{Pharma AUG2011-Paper SP06} 2011; .

\bibitem{cox2007ggm}
Cox C, Chu H, Schneider MF et~al.
\newblock Parametric survival analysis and taxonomy of hazard functions for the generalized gamma distribution.
\newblock \emph{Statistics in Medicine} 2007; 26(23): 4352--4374.
\newblock \doi{https://doi.org/10.1002/sim.2836}.
\newblock \urlprefix\url{https://onlinelibrary.wiley.com/doi/abs/10.1002/sim.2836}.
\newblock \eprint{https://onlinelibrary.wiley.com/doi/pdf/10.1002/sim.2836}.

\bibitem{Shen2022nph}
Shen YL, Wang X, Sirisha M et~al.
\newblock Nonproportional hazards—an evaluation of the maxcombo test in cancer clinical trials.
\newblock \emph{Statistics in Biopharmaceutical Research} 2022; 0(0): 1--10.
\newblock \doi{10.1080/19466315.2021.2008485}.
\newblock \urlprefix\url{https://doi.org/10.1080/19466315.2021.2008485}.
\newblock \eprint{https://doi.org/10.1080/19466315.2021.2008485}.

\bibitem{royston2020simulation}
Royston P and B~Parmar MK.
\newblock A simulation study comparing the power of nine tests of the treatment effect in randomized controlled trials with a time-to-event outcome.
\newblock \emph{Trials} 2020; 21(1): 1--17.

\bibitem{cox2008gfm}
Cox C.
\newblock The generalized f distribution: an umbrella for parametric survival analysis.
\newblock \emph{Statistics in medicine} 2008; 27(21): 4301--4312.

\bibitem{GTtest}
GRAMBSCH PM and THERNEAU TM.
\newblock Proportional hazards tests and diagnostics based on weighted residuals.
\newblock \emph{Biometrika} 1994; 81(3): 515--526.
\newblock \doi{10.1093/biomet/81.3.515}.
\newblock \urlprefix\url{https://doi.org/10.1093/biomet/81.3.515}.
\newblock \eprint{https://academic.oup.com/biomet/article-pdf/81/3/515/714158/81-3-515.pdf}.

\bibitem{xue2020online}
Xue Y, Wang H, Yan J et~al.
\newblock An online updating approach for testing the proportional hazards assumption with streams of survival data.
\newblock \emph{Biometrics} 2020; 76(1): 171--182.

\bibitem{schoenfeld1980chi}
Schoenfeld D.
\newblock Chi-squared goodness-of-fit tests for the proportional hazards regression model.
\newblock \emph{Biometrika} 1980; 67(1): 145--153.

\bibitem{abeysekera2009use}
Abeysekera W and Sooriyarachchi M.
\newblock Use of schoenfeld’s global test to test the proportional hazards assumption in the cox proportional hazards model: an application to a clinical study.
\newblock \emph{JNatnSciFoundation Sri Lanka} 2009; 37(1): 41--45.

\bibitem{fleming2011counting}
Fleming TR and Harrington DP.
\newblock \emph{Counting processes and survival analysis}.
\newblock John Wiley \& Sons, 2011.

\bibitem{andersen2004regression}
Andersen PK, Hansen MG and Klein JP.
\newblock Regression analysis of restricted mean survival time based on pseudo-observations.
\newblock \emph{Lifetime data analysis} 2004; 10: 335--350.

\bibitem{Royston2011RMST}
Royston P and Parmar MKB.
\newblock The use of restricted mean survival time to estimate the treatment effect in randomized clinical trials when the proportional hazards assumption is in doubt.
\newblock \emph{Statistics in Medicine} 2011; 30(19): 2409--2421.
\newblock \doi{https://doi.org/10.1002/sim.4274}.
\newblock \urlprefix\url{https://onlinelibrary.wiley.com/doi/abs/10.1002/sim.4274}.

\bibitem{royston2013restricted}
Royston P and Parmar MK.
\newblock Restricted mean survival time: an alternative to the hazard ratio for the design and analysis of randomized trials with a time-to-event outcome.
\newblock \emph{BMC medical research methodology} 2013; 13(1): 1--15.

\bibitem{TrinquartRMSTratio}
Trinquart L, Jacot J, Conner SC et~al.
\newblock Comparison of treatment effects measured by the hazard ratio and by the ratio of restricted mean survival times in oncology randomized controlled trials.
\newblock \emph{Journal of Clinical Oncology} 2016; 34(15): 1813--1819.
\newblock \doi{10.1200/JCO.2015.64.2488}.
\newblock \urlprefix\url{https://doi.org/10.1200/JCO.2015.64.2488}.
\newblock PMID: 26884584.

\bibitem{Tianselecttimewindow}
Tian L, Jin H, Uno H et~al.
\newblock On the empirical choice of the time window for restricted mean survival time.
\newblock \emph{Biometrics} 2020; 76(4): 1157--1166.
\newblock \doi{https://doi.org/10.1111/biom.13237}.
\newblock \urlprefix\url{https://onlinelibrary.wiley.com/doi/abs/10.1111/biom.13237}.
\newblock \eprint{https://onlinelibrary.wiley.com/doi/pdf/10.1111/biom.13237}.

\bibitem{jackson2010survival}
Jackson CH, Sharples LD and Thompson SG.
\newblock Survival models in health economic evaluations: balancing fit and parsimony to improve prediction.
\newblock \emph{The international journal of biostatistics} 2010; 6(1).

\bibitem{box2004event}
Box-Steffensmeier JM and Jones BS.
\newblock \emph{Event history modeling: A guide for social scientists}.
\newblock Cambridge University Press, 2004.

\bibitem{jackson2016flexsurv}
Jackson CH.
\newblock flexsurv: a platform for parametric survival modeling in r.
\newblock \emph{Journal of statistical software} 2016; 70.

\bibitem{wei1992accelerated}
Wei LJ.
\newblock The accelerated failure time model: a useful alternative to the cox regression model in survival analysis.
\newblock \emph{Statistics in medicine} 1992; 11(14-15): 1871--1879.

\bibitem{prentice1975discrimination}
Prentice RL.
\newblock Discrimination among some parametric models.
\newblock \emph{Biometrika} 1975; 62(3): 607--614.

\bibitem{lin2020alternative}
Lin RS, Lin J, Roychoudhury S et~al.
\newblock Alternative analysis methods for time to event endpoints under nonproportional hazards: a comparative analysis.
\newblock \emph{Statistics in Biopharmaceutical Research} 2020; 12(2): 187--198.

\bibitem{Guyot2012dataconstruct}
P G, A A and et~al OM.
\newblock Enhanced secondary analysis of survival data: reconstructing the data from published kaplan-meier survival curves.
\newblock \emph{BMC Med Res Methodol} 2012; 12(9).
\newblock \urlprefix\url{https://doi.org/10.1186/1471-2288-12-9}.

\bibitem{benboubker2014lenalidomide}
Benboubker L, Dimopoulos MA, Dispenzieri A et~al.
\newblock Lenalidomide and dexamethasone in transplant-ineligible patients with myeloma.
\newblock \emph{New England Journal of Medicine} 2014; 371(10): 906--917.

\bibitem{kantarjian2016inotuzumab}
Kantarjian HM, DeAngelo DJ, Stelljes M et~al.
\newblock Inotuzumab ozogamicin versus standard therapy for acute lymphoblastic leukemia.
\newblock \emph{New England Journal of Medicine} 2016; 375(8): 740--753.

\bibitem{burger2011incorporation}
Burger RA, Brady MF, Bookman MA et~al.
\newblock Incorporation of bevacizumab in the primary treatment of ovarian cancer.
\newblock \emph{New England Journal of Medicine} 2011; 365(26): 2473--2483.

\bibitem{GOG-0218info}
Bevacizumab-ProductLabel.
\newblock Genentech.
\newblock \url{https://www. accessdata.fda.gov/drugsatfda_docs/label/2020/125085s337lbl.pdf }, 2020.

\bibitem{Ray2020}
Lin RS, Lin J, Roychoudhury S et~al.
\newblock Alternative analysis methods for time to event endpoints under nonproportional hazards: A comparative analysis.
\newblock \emph{Statistics in Biopharmaceutical Research} 2020; 12(2): 187--198.
\newblock \doi{10.1080/19466315.2019.1697738}.
\newblock \urlprefix\url{https://doi.org/10.1080/19466315.2019.1697738}.
\newblock \eprint{https://doi.org/10.1080/19466315.2019.1697738}.

\bibitem{rajkumar2010lenalidomide}
Rajkumar SV, Jacobus S, Callander NS et~al.
\newblock Lenalidomide plus high-dose dexamethasone versus lenalidomide plus low-dose dexamethasone as initial therapy for newly diagnosed multiple myeloma: an open-label randomised controlled trial.
\newblock \emph{The lancet oncology} 2010; 11(1): 29--37.

\bibitem{nsclc2008chemotherapy}
Group NMAC et~al.
\newblock Chemotherapy in addition to supportive care improves survival in advanced non--small-cell lung cancer: a systematic review and meta-analysis of individual patient data from 16 randomized controlled trials.
\newblock \emph{Journal of Clinical Oncology} 2008; 26(28): 4617.

\end{thebibliography}

\newpage
\section*{Supplementary material}

\setcounter{figure}{0}
\setcounter{table}{0}
\renewcommand{\thefigure}{S\arabic{figure}}
\renewcommand{\thetable}{S\arabic{table}}

\begin{figure}[ht]
    \centering
    \subfigure[Kaplan–Meier curves for progression-free survival: Study FIRST (Revlimid).]{
    \includegraphics[width=2.5in]{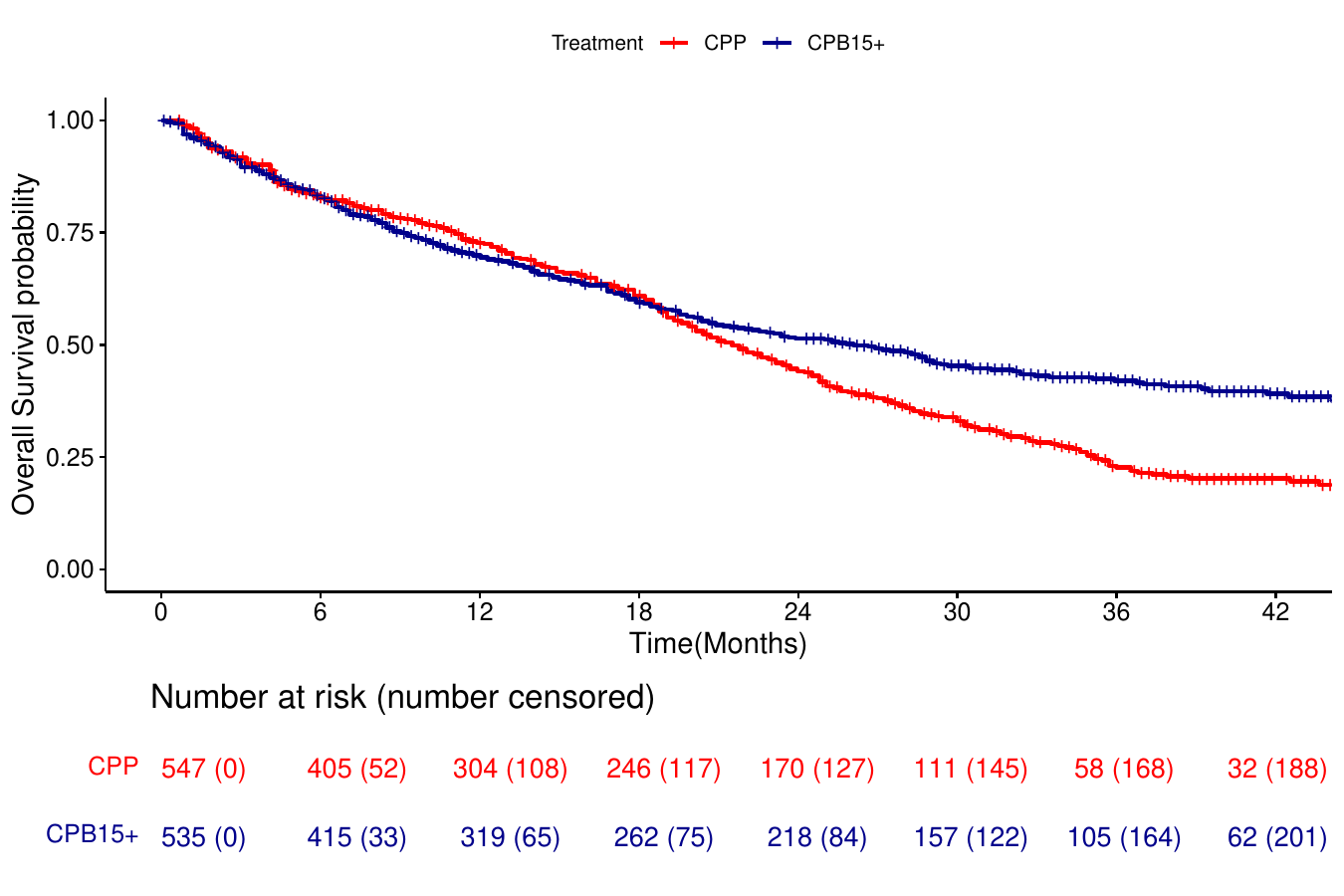}
    \label{fig:study1_ori}
    }
    \subfigure[Kaplan–Meier curves for overall survival: Study INO-VATE (Inotuzumab).]{
    \includegraphics[width=2.5in]{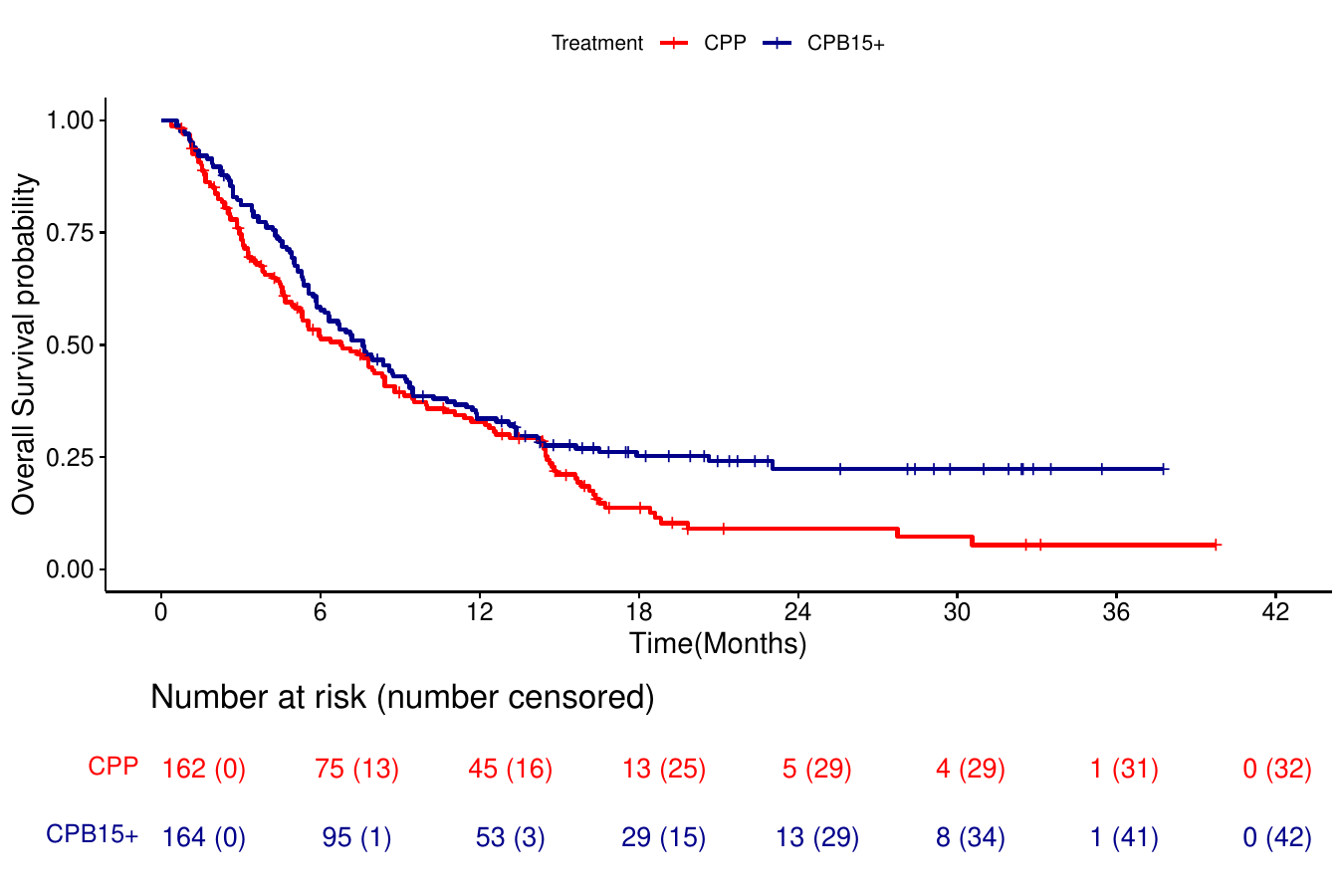}
    \label{fig:study2_ori}
    }
    \subfigure[Kaplan–Meier curves for progression-free survival: Study GOG-0218 (Bevacizumab).]{
    \includegraphics[width=2.5in]{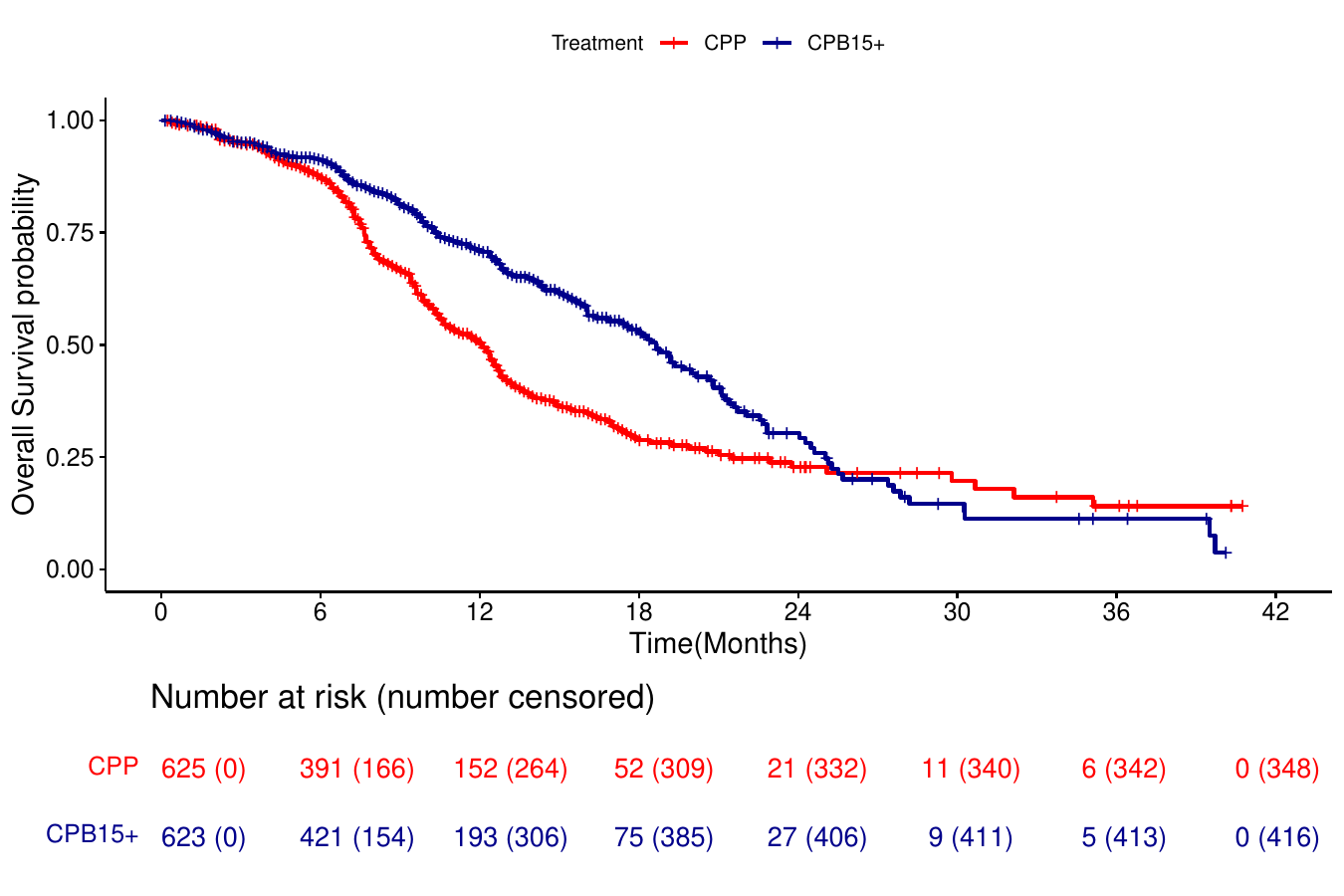}
    \label{fig:study3_ori}
    }
    
\caption{Kaplan–Meier curves of three published oncology studies.}
\label{fig:study_ori}
\end{figure}

\end{document}